\documentclass{article}
\usepackage{amsmath}
\usepackage{amsfonts}

\usepackage[space]{cite}
\def\@biblabel#1{[#1]}

\usepackage{hyperref}

\newcommand{\tr}{\mathop{\rm tr}\nolimits}

\title{Entanglement in Fermionic Chains and Bispectrality}

\author{Nicolas Cramp\'e\thanks{Institut Denis-Poisson CNRS/UMR 7013 - Universit\'e de Tours - Universit\'e d'Orl\'eans, 
Parc de Grandmont, 37200 Tours, France  {\tt crampe1977@gmail.com}},  
Rafael I. Nepomechie\thanks{Physics Department, P.O. Box 248046, University of Miami, 
Coral Gables, FL 33124 USA {\tt nepomechie@miami.edu}} 
and Luc Vinet\thanks{Centre de Recherches Math\'ematiques, Universit\'e de Montr\'eal,
P.O. Box 6128, Centre-ville Station, Montr\'eal (Qu\'ebec), H3C 3J7, 
Canada {\tt vinet@crm.umontreal.ca}}}

\date{}

\begin{document}

\maketitle

\begin{abstract}
	Entanglement in finite and semi-infinite free Fermionic chains is
	studied.  A parallel is drawn with the analysis of time and band
	limiting in signal processing.  It is shown that a tridiagonal
	matrix commuting with the entanglement Hamiltonian can be found
	using the algebraic Heun operator construct in instances when
	there is an underlying bispectral problem.  Cases corresponding to
	the Lie algebras $\mathfrak{su}(2)$ and $\mathfrak{su}(1,1)$ as
	well as to the q-deformed algebra $\mathfrak{so}_q(3)$ at $q$ a
	root of unity are presented.
\end{abstract}

\bigskip

\textit{This paper is dedicated to Roman Jackiw with admiration and
gratitude on the occasion of his 80th birthday.}

\setcounter{footnote}{0}

\section{Introduction}
Throughout his career Roman Jackiw has achieved a number of important
scientific advances and in the process he has brought many modern
geometrical, topological and representation theoretic results to bear
on the elaboration and understanding of physical theories.  He has
hence much contributed to increasing the level of interactions between
physicists and mathematicians.  We here wish to thankfully pay tribute
to him by discussing how symmetry and algebraic considerations can
contribute to entanglement studies in light of a parallel with
long-studied issues in signal processing.  We hope that this report
will hence capture some of the bridge building
spirit of Roman's insightful and inspiring papers.

A fundamental feature of quantum theories, entanglement enables
correlations and is a key resource in applications to information.  It
is therefore relevant to obtain quantitative evaluation of this
property and this is being much explored using the notion of entropy.
This paper belongs to that class of studies and focuses on systems
consisting of free Fermionic chains that have been much looked at
\cite{2004JSMTE..06..004P, 2018arXiv180500078E}, because of their
simplicity.

Basically, for these systems, the entanglement entropy is determined
by the eigenvalues of a truncated correlation matrix.  However a
significant difficulty in carrying their computation arises for large
chains because the spectra of these correlation matrices usually
accumulates near certain points thereby rendering the numerical
analysis problematic.  As a matter of fact this entanglement problem
proves closely analogous \cite{2006PhRvL..96j0503G, 2004JSMTE..06..004P, 2013JSMTE..04..028E,
2018arXiv180500078E} to the classical question of the time and
band limiting of signals where the corresponding calculation
difficulty is circumvented thanks to the discovery made by Slepian et
al.  \cite{MR710468, Landau1985, Slepian1961} of an operator easy to handle
numerically that commutes with the limiting operator.

The main goal of the present paper is to explain how this efficient
processing of signals can be adapted in certain cases to the entropy
analysis of Fermionic chains.

Although this is still not fully understood, the circumstances for the
existence of the commuting operator are perceived to stem from
bispectrality situations \cite{1986CMaPh.103..177D, doi:10.1002/cpa.3160470305} where the
functions involved depend on two variables and satisfy a pair of
eigenvalue problems such that in the first equation the operator acts
on one variable and the eigenvalue is a function of the other variable
and vice-versa in the second equation.  The hypergeometric polynomials
of the Askey scheme \cite{Koekoek, Koekoek2010}, offer examples
of such bispectral problems: they are eigenfunctions of a
differential or difference operator in the variable with the
eigenvalues depending on the degree and, their orthogonality requires
that they satisfy a recurrence relation which is viewed as an
eigenvalue equation for an operator acting on the degree taken as a
discrete variable with the eigenvalue in this case solely depending on
the standard variable.  One notes that there are two pictures for the
pair of operators: the variable picture and the degree picture much
like the coordinate and momentum representations in Quantum Mechanics.

That bispectrality has something to do with the existence of a
commuting operator in problems of the time and band limiting class was
revisited recently in Ref. \cite{2017JMP....58c1703G}.  Assume that
the limiting takes place by restricting the range of the two variables
associated to the problem.  A first observation is that the bilinear
combinations of the two bispectral operators provide generalizations
of the Heun operator which itself actually arises in the particular
case of the Jacobi polynomials.  The reader will recall that the usual
Heun operator defines the Fuchsian second order differential equation
with four regular singularities.  To each bispectral problem is thus
associated what has been called an algebraic Heun operator.  Once this
is recognized, it is easy to determine how these generalized Heun
operators should be specialized so as to commute with the projectors
on the restrained domains and as a consequence with the limiting
operator.

Basically, determining the entanglement of Fermionic chains amounts:
i.  to taking the chain in some state which we will assume to be the
ground state, ii.  to dividing the system into two parts, and iii.  to
examining how these two parts are coupled in the chosen state.  The
analogy with time and band limiting arises as energy is gapped by the
Fermi sea filling and space is chopped through the partitioning of the
chain.  In cases where the Fermionic chain Hamiltonian exhibits
bispectral features, we shall show how algebraic Heun operators
provide matrices that commute with the entanglement Hamiltonian and
have nice properties from the point of view of numerical analysis.

This paper enlarges and complements our recent article
\cite{Cramp__2019} on this topic where the emphasis in the
characterization of the chains and their properties was put on the
associated orthogonal polynomials.  Here the focus is on bispectrality
and algebras.  The parallel with time and band limiting will be
explained with the help of a review of the classic results in this
field and the connection with algebraic Heun operators will be
illustrated in this context first.  Supplementing the set of chains
considered in \cite{Cramp__2019}, we shall discuss a
semi-infinite chain as well as a finite one based on a representation
of a $q$-deformed algebra at $q$ a root of unity that has as special
case the uniform chain treated in \cite{2018arXiv180500078E} and
\cite{Cramp__2019}.

The presentation will proceed along the following lines.  The free
Fermionic Hamiltonians and their diagonalization are described in
Section 2 that will also establish notation.  Section 3 introduces the
restricted correlation matrix as the central quantity for the study of
entanglement.  Section 4 momentarily leaves the topic of Fermionic
chains to offer a short overview of the classical problem of limiting in
time a signal which is banded in frequency.  It shall explain how the
Heun operator associated to the Fourier bispectral problem leads to
the second order differential operator that commutes with the integral
operator that effects the limiting in this case.  Section 5 returns to
Fermionic chains in light of this understanding and discusses
generally when the Hamiltonians are characterized by a bispectral
problem.  For finite chains this will involve Leonard pairs which are
known to be in correspondence with the families of orthogonal
polynomials of the terminating branch of the Askey scheme.  Section 6
derives the tridiagonal matrices that commute with the chopped
correlation matrix from the algebraic Heun operators attached to
Hamiltonians with bispectral underpinnings.  Special bispectral
situations that will be considered as examples shall be arising from
the representation theory of Lie and $q$-algebras.  This will be
the contents of Sections 7, 8, and 9.  Section 7 will reproduce
results from \cite{Cramp__2019} by discussing the chain based on
$\mathfrak{su}(2)$.  Section 8 will treat the case of the
semi-infinite chain associated to $\mathfrak{su}(1,1)$.  Section 9
will focus on the chain whose couplings are given by the
representation matrices of the non-standard deformation
$\mathfrak{so}_q(3)$ of $\mathfrak{so}(3)$ at $q$ root of unity.  This
will have as a special case the uniform chain treated in
Refs. \cite{2018arXiv180500078E} and \cite{Cramp__2019}.  Section 10 shall bring
the paper to a close with concluding remarks.

\section{Free-Fermion Hamiltonian and its diagonalization}

We consider the following open quadratic free-Fermion inhomogeneous
Hamiltonian with nearest neighbour interactions and with magnetic 
fields
\begin{equation}\label{eq:Hff}
\widehat{\mathcal{H}}=\sum_{n=0}^{N-1}(J_{n}  c_n^\dagger c_{n+1} + J_n^* c_{n+1}^{\dagger} c_{n})
- \sum_{n=0}^{N}B_n c_{n}^{\dagger} c_{n},
\end{equation}
where $B_n$ (resp.  $J_n$) are real (resp.  complex) parameters,
$J_n^*$ is the complex conjugate of $J_n$ and $\{ c_{m}^{\dagger}
\,, c_{n} \} = \delta_{m,n}$. 
For the sake of simplicity of the following computations, we enumerate
the sites of the lattice from $0$ to $N$.  We can also consider the
case $N\rightarrow +\infty$ which corresponds to a
semi-infinite chain (see Section \ref{sec:su11} for
an example).

In order to diagonalize $\widehat{\mathcal{H}}$, 
it is convenient to rewrite it as follows
\begin{equation}\label{eq:Hff2}
\widehat{\mathcal{H}}=(c^\dagger_0,\dots,c^\dagger_N)\,  \widehat H \begin{pmatrix}c_0\\ \vdots \\c_N\end{pmatrix} \ .
\end{equation}
The $(N+1)\times (N+1)$ matrix $\widehat H$ is an Hermitian tridiagonal matrix given by
\begin{equation}\label{eq:Hh}
 \widehat H  =\sum_{n=0}^N\Big(J_{n-1} |n-1\rangle \langle n|  -B_n |n\rangle\langle n|  + J_{n}^*
 |n+1\rangle\langle n|\Big)\,, 
\end{equation}
with the convention $J_{N}=J_{-1}=0$. The set $\{|0\rangle,|1\rangle,\dots ,|N\rangle$ 
of elements in $\mathbb{C}^{N+1}$ denotes the canonical orthonormal basis and will be called
the position basis.
The spectral problem for $\widehat H$ reads
\begin{equation}
 \widehat H |\omega_k\rangle = \omega_k |\omega_k\rangle\ ,
 \label{eigenh}
\end{equation}
where
\begin{equation}
 |\omega_k\rangle =\sum_{n=0}^N \phi_n(\omega_k) |n\rangle\ . \label{eq:omm}
\end{equation}
We order the $N+1$ eigenvalues $\omega_0, \omega_1, \dots 
\omega_N$ so that $\omega_k < \omega_{k+1}$.
We also normalize the eigenvectors  $|\omega_0\rangle,|\omega_1\rangle, \dots 
|\omega_N\rangle$ so that they form an orthonormal basis of 
$\mathbb{C}^{N+1}$, to be called the momentum basis. 
Having diagonalized $\widehat H$, we see that the Hamiltonian 
$\widehat{\mathcal{H}}$ \eqref{eq:Hff} can be rewritten as 
\begin{equation}
\widehat{\mathcal{H}}=\sum_{k=0}^{N} \omega_{k} 
\tilde{c}^{\dagger}_{k} \tilde{c}_{k} \,, 
\end{equation}
where the annihilation operators $\tilde{c}_{k}$ are defined by
\begin{equation}
\tilde{c}_{k} = \sum_{n=0}^{N} \phi_{n}^*(\omega_{k})\, 
c_{n} \,, \qquad
\label{ctilde}
\end{equation}
and the corresponding formulas for the creation operators  
$\tilde{c}^{\dagger}_{k}$ are given by the 
Hermitian conjugation of \eqref{ctilde}.
These operators obey the anticommutation relations
\begin{equation}
	\{ \tilde{c}^{\dagger}_{k} \,, \tilde{c}_{p} \} = \delta_{k,p} 
	\,, \qquad  
\{ \tilde{c}^{\dagger}_{k} \,, \tilde{c}^{\dagger}_{p} \} = \{ \tilde{c}_{k} \,, 
\tilde{c}_{p} \} = 0 \,.
\label{CR}
\end{equation}
One can invert relation \eqref{ctilde} to get
\begin{equation}
c_{n} = \sum_{k=0}^{N}\phi_{n}(\omega_{k})\, \tilde{c}_{k} \,.
\label{cctilde}
\end{equation}
The eigenvectors of $\widehat{\mathcal{H}}$ are therefore given by
\begin{equation}
	|\Psi\rangle\!\rangle = \tilde{c}_{k_{1}}^\dagger \ldots \tilde{c}_{k_{r}}^\dagger |0\rangle\!\rangle \,,
	\label{grounstate}
\end{equation}
where $k_{1}< \ldots < k_{r} \in \{0, \ldots, N\}$, and
the vacuum state $|0\rangle\!\rangle$ is annihilated by all the annihilation 
operators
\begin{equation}
	\tilde{c}_{k}|0\rangle\!\rangle = 0\,, \qquad k = 0\,, \ldots\,, 
	N \,.
\label{vacuum}
\end{equation}
	The corresponding energy eigenvalues of $\widehat{\mathcal{H}}$ are simply given by
\begin{equation}
	E = \sum_{i=1}^{r} \omega_{k_{i}} \,.
\end{equation}

\section{Correlations and the entanglement Hamiltonian}

For the sake of concreteness, we shall consider entanglement in the
ground state described below.  We shall further review how
the reduced density matrix for the first $\ell+1$ sites of the
chain is determined by the 1-particle correlation matrix and 
equivalently by the entanglement Hamiltonian.  

\subsection{Defining the ground state}\label{subsec:chop}

The fact that the ground state is constructed by filling the Fermi sea leads to a
restriction in energy. Indeed, the ground state $|\Psi_{0}\rangle\!\rangle$ of 
the Hamiltonian \eqref{eq:Hff} is given by
\begin{equation}
	|\Psi_{0}\rangle\!\rangle = \tilde{c}_0^\dagger \ldots \tilde{c}_K^\dagger |0\rangle\!\rangle \,,
\end{equation}
where $K\in \{0,1,\dots,N\}$ is the greatest integer below the Fermi
momentum, such that
\begin{equation}
	\omega_{K} < 0\,, \qquad \omega_{K+1} > 0 \,.
	\label{Kdef}
\end{equation}	
Let us remark that $K$ can be modified by adding a constant term to
the external magnetic fields $B_n$.  We shall in fact choose
this constant magnetic field so as to ensure that $\omega_k\neq 0$ 
for any $k$ in order to avoid dealing with a degenerate ground state.

The correlation matrix $\widehat C$ in the ground state is an $(N+1)
\times (N+1)$ matrix with the following entries
\begin{equation}
\widehat C_{mn}=\langle\!\langle \Psi_{0}| c_m^\dagger c_n |\Psi_{0} \rangle\!\rangle \,.
\label{Cmatrixdef}
\end{equation}
Expressing everything in terms of annihilation and creation operators
using \eqref{cctilde} and \eqref{grounstate}, and then using the
anticommutation relations \eqref{CR} and the property \eqref{vacuum} 
of the vacuum state, we obtain
\begin{equation}
\widehat C_{mn}=\sum_{k=0}^K 
\phi_m^*(\omega_k) \phi_n(\omega_k)\,, \qquad 0\leq n,m\leq N \,.
\label{Cmatrix}
\end{equation}
It is then manifest that 
\begin{equation}
 \widehat C = \sum_{k=0}^K  |\omega_k\rangle \langle \omega_k| \,,
 \label{Cdag}
\end{equation}
namely, that $\widehat C$ is the projector onto the subspace of
$\mathbb{C}^{N+1}$ spanned by the vectors $|\omega_k\rangle$ with
$k=0,...,K$ running over the labels of the excitations in the ground
state.

\subsection{Entanglement entropy}

In order to examine entanglement, we must first define a bipartition
of our free-Fermionic chain. As
subsystem (part 1) we shall take the first $\ell+1$ consecutive sites, and shall
find how it is intertwined with the rest of the chain in
the ground state $|\Psi_{0}\rangle\!\rangle$.  To that end, we need the
reduced density matrix 
\begin{equation}
\rho_{1} = \tr_{2} |\Psi_{0}\rangle\!\rangle 
\langle\!\langle \Psi_{0}| \,,
\end{equation}
where part $2$, the complement of part $1$, is comprised of the sites
$\{ \ell+1, \ell+2,...,N\}$; from this quantity one can compute for instance the von
Neumann entropy 
\begin{equation}
S_{1} = -\tr
(\rho_{1} \log \rho_{1})\ .\end{equation}
The explicit computations of this entanglement entropy amounts to finding the eigenvalues of
$\rho_{1}$.

It has been observed that this reduced density matrix $\rho_1$ is determined by
the spatially ``chopped'' correlation matrix $C$, 
which is the following $(\ell+1) \times (\ell+1)$ submatrix of $\widehat C$: 
\begin{equation}
  C =  |\widehat C_{mn}|_{0\leq m,n \leq \ell}\,. 
\end{equation}
The argument which we take from Ref. \cite{2003JPhA...36L.205P} (see also
Ref. \cite{2009JPhA...42X4003P}) goes as follows.  Because the ground state
of the Hamiltonian $\widehat{\mathcal{H}}$ is a Slater determinant,
all correlations can be expressed in terms of the one-particle
functions, i.e. in terms of the matrix elements of $\widehat C$. 
Restricting to observables $A$ associated to part 1, since the
expectation value of $A$ is given by $\langle A \rangle = \tr (\rho_{1}A)$,
the factorization property will hold according to Wick's theorem if 
$\rho_{1}$ is of the form 
\begin{equation}
\rho_{1} = \kappa \; \exp (-\mathcal{H}) \,,
\label{entH}
\end{equation}
with the entanglement Hamiltonian $\mathcal{H}$ given by
\begin{equation}
\mathcal{H} = \sum_{m,n =0}^\ell h_{mn} \, c_{m}^\dagger 
c_{n} \,.
\label{hopping}
\end{equation}

The hopping matrix $h= |h_{mn}|_{0 \le m,n \le \ell}$ is defined so that
\begin{equation}
C_{mn} = \tr (\rho_{1} \; c_m^\dagger c_n)\,,  \qquad  m\,, n \in \{0, 
1,\dots , \ell\},
\label{obs}
\end{equation}
holds, and one 
finds through diagonalization that 
\begin{equation}
h = \log [(1 - C) /C] \,.
\end{equation}
We thus see that the $2^{(\ell+1)} \times 2^{(\ell+1)}$ matrix $\rho_{1}$
is obtained from the $(\ell+1) \times (\ell+1)$ matrix $C$ or
equivalently, from the entanglement Hamiltonian $\mathcal{H}$.

Introducing the projectors 
\begin{equation}
 \pi_1=\sum_{n=0}^\ell  |n\rangle \langle n| \quad \text{and}\quad  
 \pi_2=\sum_{k=0}^K   |\omega_k\rangle \langle \omega_k| = \widehat C \,,
 \label{proj}
\end{equation}
the chopped correlation matrix can be written as (see for instance 
Refs. \cite{Lee:2014nra, 2012PhRvB..86x5109H})
\begin{equation}
 C =\pi_1 \pi_2 \pi_1 \ .
 \label{Cpi}
\end{equation}

To calculate the entanglement entropies one therefore has to compute the
eigenvalues of $C$.  As explained in Ref. \cite{2004JSMTE..06..004P}, this
is not easy to do numerically because the eigenvalues of that matrix
are exponentially close to $0$ and $1$. We shall show in the following how to go about
this problem by drawing on methods developed in signal processing.

\section{A review of time and band limiting \label{sec:BTL}}

We here digress to underscore that the treatment of time and band
limiting problems is of relevance for the characterization of
entanglement in Fermionic chains.  To make that clear, we shall review
aspects of the classic problem of optimizing the concentration in time
of a band limited signal.  In the first part of this section we shall
show that the limiting integral operator can also be expressed in
terms of projectors exactly as in \eqref{Cpi}.  The diagonalization of
this operator that would give the optimization solution is also
plagued by computational difficulties.  In the second part of the
section we shall indicate how the underlying bispectrality provides a
way to overcome this numerical analysis challenge by allowing to
identify a differential operator that commutes with the limiting one.

Let $f(t)$ be a signal limited to the band of frequencies $[-W,W]$: 
\begin{equation}
f(t)=\frac{1}{\sqrt{2\pi}}\int_{-W}^{W}dp \;e^{ipt} F(p) \in B_W\, ,
\end{equation}
and call $B_W$ the space of such functions taken to be real. 
It is natural to want a signal of finite duration, that is to ask that $f(t)$ vanishes outside the interval $ -T<t<T$: 
\begin{equation}
f\neq 0 \quad \text{only for}\quad  -T<t<T.
\end{equation}
It is however readily realized that this is impossible: since $f(t) 
\in B_W$, it is entire in complex $t$-plane; therefore
if $f(t)=0$ for any interval, it follows that $f(t)$ is identically zero
($f(t) \equiv 0$). In the 1960s and 1970s Slepian, Landau, Pollak 
from Bell labs (see the reviews \cite{MR710468, Landau1985}) 
considered how to approximate the situation wished for and asked the question:
Which band-limited signal $\in B_W$ is best concentrated in the time interval $-T<t<T$, second best concentrated etc.? In other words which 
functions $f(t) \in B_W$ are maximizing 
\begin{align}
 \alpha^2(T)&=\frac{\displaystyle \int_{-T}^{T}f^2(t)dt}{\displaystyle \int_{-\infty}^{\infty}f^2(t)dt}\\
&=2 \ 
 \frac{\displaystyle 
 \int_{-W}^{W}dp' \int_{-W}^{W}dp'' 
 \left[  \frac{\sin((p'-p'')T)}{(p'-p'')}\right ] 
 F(p'')F^*(p')}
 {\displaystyle \int_{-W}^{W} dp' F(p')F^*(p')}.
 \end{align}
As is well known from the calculus of variations, the answer to that question is provided by the solutions of 
\begin{equation}
GF(p)=\lambda F(p)\, ,
\end{equation}
where the integral operator $G$ is defined by
\begin{equation}
GF(p)=\int_{-W}^{W}dp'K(p-p')F(p') \, ,
\end{equation}
with $K(p-p')$ the sinc kernel 
\begin{equation}
K(p-p')=\frac{\sin((p-p')T)}{\pi(p-p')}\ .
\end{equation} 
Let us remark that $GF(p)$ is zero if $p$ is not between $-W$ and $W$ as the functions $F(p)$ we start with.

In principle this should settle the concentration problem.  However
the spectrum of $G$ accumulates sharply at the origin and this makes
the numerical computations intractable.  Slepian, Landau, Pollak 
\cite{MR710468, Landau1985, Slepian1961}
quite remarkably found a way out by showing that there exists a second order
differential operator $D$ that commutes with the integral operator
$G$.  This is important because $D$ has common eigenfunctions with $G$ and
second order differential operator are typically well behaved
numerically.  It is interesting to mention that $D$ actually arises in
separating the Laplacian in prolate spheroidal coordinates.  Let us
indicate how this commuting operator is obtained using the bispectral
framework of Fourier transform.

We shall first note that $G$ can be written in a form similar to that
given in \eqref{Cpi} for the chopped correlation matrix.  Consider
projectors on an interval:
\begin{eqnarray}
\pi^x_Lf(x)&=&\begin{cases} f(x) \quad -L<x<L\\ 0 \quad \text{otherwise}\end{cases}\\
 &=&[\Theta(x+L)-\Theta(x-L)]f(x)\,,
\end{eqnarray}
with $\Theta(x) $ the step function.
Let $\mathcal{F} :  f(t) \mapsto F(p)$ denote the Fourier transform and $\mathcal{F}^{-1}$ its inverse.
Take the following projectors in Fourier (or band) space:
\begin{equation}
\pi^p_W \qquad \text{and the Fourier transformed} \qquad \hat{\pi}^{p}_T = \mathcal{F} \pi^t_T \mathcal{F}^{-1}\,.
\end{equation}
 It is straightforward to see that 
 \begin{equation}
 G=\pi^p_W \hat \pi^p_T \pi^p_W \,.
 \label{G}
 \end{equation}

Operators $X$ and $Y$ form a bispectral pair if 
they have common eigenfunctions $\psi(x,n)$ such that 
\begin{align}
 X \psi(x,n) &= \omega(x) \psi(x,n) \,, \\
 Y \psi(x,n) &= \lambda(n) \psi(x,n) \,,
\end{align}
with $X$ acting on the variable $n$ and $Y$, on the variable $x$. 
When forming products of these operators $X$ and $Y$, we shall understand that they are both taken in same representation ``$n$'' or ``$x$''. 
The functions $\psi(t, p)=e^{ipt}$ in Fourier transforms satisfy
\begin{equation}
-\frac{d^2}{dt^2}\psi(t,p)=p^2 \psi(t,p), \qquad -\frac{d^2}{dp^2}\psi(t,p)=t^2\psi(t,p) \,, \label{eq:bis}
\end{equation}
and are thus associated to a most simple bispectral problem: the
functions $\psi(t, p)$ are eigenfunctions of an operator acting on $t$
with eigenvalues depending on $p$ and vice-versa.  All the orthogonal
polynomials of the Askey scheme are solutions of bispectral problems
defined by the differential/difference equation and the recurrence
relation.

How does this help find the differential operator that commutes with the limiting operator $G$?

To each bispectral problem, one can attach an \textit{Algebraic Heun Operator} \cite{2018CMaPh.364.1041G} defined as 
the most general operator $W_H$ which is bilinear in the bispectral operators $X$ and $Y$:
\begin{equation}\label{eq:heun2}
W_H= \tau_1 \{X,Y\} + \tau_2 [X,Y] + \tau_3 X + \tau_4 Y + \tau_0  \,, 
\end{equation}
with $\tau_i, i= 0, 1\dots,4$, constants and $\{X,Y\}=XY+YX$.  The
name comes from the fact that the standard Heun operator results when
this construct is applied to the bispectral operators of the Jacobi
polynomials, namely the hypergeometric operator and multiplication by
the variable $x$.  We claim that the commuting operators belong to
that class of operators.  Let us return to the Fourier case where in
the ``frequency'' representation
\begin{equation}
X=-\frac{d^2}{dp^2}, \qquad Y=p^2.
\end{equation}
In this representation, taking $\tau_0=0$ and $\tau_1=-1/2$, the algebraic Heun operator which we will now denote by $D$ takes the form:
\begin{align}
D &= \frac{1}{2}\{ \frac{d^2}{dp^2}, p^2 \} + \tau [-\frac{d^2}{dp^2}, p^2] - \mu \frac{d^2}{dp^2}+\nu p^2\\
  &= (p^2-\mu)\frac{d^2}{dp^2} + (2-4\tau)p\frac{d}{dp} + \nu p^2 - 2\tau +1.   \label{D}
\end{align}
Given \eqref{G}, such an operator will commute with $G$ if it commutes with both $\pi^p_W$ and $ \hat \pi^p_T$.
Consider a general second order differential operator written as
\begin{equation}
\mathcal{D}= A(p)\frac{d^2}{dp^2} + B(p) \frac{d}{dp} + C(p)\,.
\end{equation}
Let us look first at the projector onto the semi-infinite interval $[W,\infty)$
\begin{equation}
\tilde \pi^p_W = \Theta (p-W)\,.
\end{equation}
It is easy to see that
$[\mathcal{D},\tilde \pi^p_W] = 2 A(p) \delta (p-W) \frac{d}{dp} + (-A'(p) +B(p)) \delta (p-W) = 0$ if 
$A(W)=0$ and $A'(W)=B(W)$. Now recall that
\begin{equation*}
\pi^p_W= \Theta(p+W) -  \Theta(p-W)\, .
\end{equation*}
In this case $[\mathcal{D},\pi^p_W]=0$ is satisfied if
\begin{equation}
A(\pm W)=0 \qquad \text{and} \qquad A'(\pm W)=B(\pm W)\, .
\end{equation}
Applying these conditions to $D$ as given by \eqref{D} is readily seen to imply that
\begin{equation}
    \mu = W^2 \qquad \text{and} \qquad \tau = 0.
\end{equation}
Now if in addition $[D,\hat \pi^p_T]=0$, we would have $[D,G]=0$. Clearly
$[D,\hat \pi^p_T]=[D,\mathcal{F}\pi^t_T\mathcal{F}^{-1}]=0$ 
is tantamount to $[\mathcal{F}^{-1}D\mathcal{F},\pi^t_T]=0$, namely to the condition that the
Fourier transform $\tilde D= \mathcal{F}^{-1}D\mathcal{F}$ of $D$ commutes with a projector in $t$ with parameter $T$ that is similar to $\pi^p_W$. Under the Fourier transform:
$ p^2 \rightarrow-\frac{d^2}{dt^2}, -\frac{d^2}{dp^2}\leftrightarrow t^2$ and $\tilde{D}$ is obtained from $D$ 
by exchanging $p$ and $t$ as well as $\mu$ and $\nu$ and by taking $\tau$ into $-\tau$. It is then obvious that the condition $[\tilde{D},\pi^t_T]=0$ is satisfied by taking
\begin{equation}
\nu = T^2 \qquad \text{and again} \qquad \tau = 0.
    \end{equation}
It thus follows that the second order differential operator that
commutes with the limiting integral operator is simply obtained from
the algebraic Heun operator \eqref{D} by imposing the conditions
$\tau =0$, $\mu = W^2$ and $\nu = T^2$ on the parameters.

The parallel with the study of entanglement in Fermionic chains is
quite clear.  Taking the chain in its ground state (or in any other
reference state) involves restricting the energies and corresponds to
band limiting.  Associated to that is the projector $\pi _2$ in
\eqref{proj}.  Establishing the bipartition truncates space and this
is akin to time limiting.  Attached to this is the projector $\pi _1$
in \eqref{proj}.  The task is to solve the eigenvalue problem for the
chopped correlation matrix $C =\pi_1 \pi_2 \pi_1$ which looks very
much like the limiting operator $G$ as given in \eqref{G} (the picture
is actually the dual one here).  On the basis of this similarity, we
may therefore hope that there could be a tridiagonal matrix - the
discrete analog of a second order differential operator - that would
commute with both $\pi _1$ and $\pi _2$ and hence with $C$ so as to
ease the numerical analysis.  Recalling that the existence of the
commuting operator was predicated on the fact that there was an
underlying bispectral problem, we shall discuss next what this
requirement entails for the specifications of the Fermionic chains
that shall henceforth be considered.

\section{A bispectral framework for Fermionic chains \label{sec:bisp}}

In order to identify Fermionic chains that are based on bispectral
problems, let us recall that two natural bases, the position basis
$\{|n\rangle\}$ and the momentum basis $\{|\omega _k\rangle\}$, are
associated to the chains.  The $(N+1)\times (N+1)$ matrix $\widehat H$
\eqref{eq:Hh} that defines the Hamiltonian is irreducible tridiagonal
in the first of these bases and diagonal in the second.  (By
irreducible it is understood that there are no zeros on the sub - and
super - diagonals.)  From \eqref{eigenh} we have
\begin{equation}
\langle n | \widehat H | \omega _k \rangle = \omega _k \langle n | \omega _k  \rangle
\end{equation}
and thus in view of \eqref{eq:Hh} the wavefunctions 
$\phi_n(\omega _k) = \langle n | \omega _k  \rangle$ satisfy the eigenvalue equation
\begin{equation}
\omega_k \phi_n(\omega_k) = J_{n}\phi_{n+1}(\omega_k) - B_n \phi_n(\omega_k) 
+  J_{n-1} \phi_{n-1}(\omega_k)\,, \qquad 
 0\leq n\leq N \,.
 \label{recurphi}
\end{equation}
We wish the functions $\phi_n(\omega _k)$ to be solutions of a
bispectral problem.  To that end we need to adjoin to $\widehat H$ a
companion operator $\widehat X$ with the property of being diagonal in
the basis $\{|n\rangle\}$ and irreducible tridiagonal in the basis
$\{|\omega _k\rangle\}$.  In other words we need an $\widehat X$ such
that
\begin{equation}
 \widehat X  = \sum_{n=0}^N \lambda_n |n\rangle\langle n |\,,
 \label{Xmat}
\end{equation}
and
\begin{equation}
 \widehat X  = \sum_{k=0}^N\Big(
 \overline{J}_{k-1} |\omega_{k-1} \rangle  \langle \omega_{k} |
 - \overline{B}_{k}|\omega_{k}\rangle\langle \omega_{k}| 
 + \overline{J}^*_{k} |\omega_{k+1}\rangle\langle \omega_{k}| \Big)  
 \,,
 \label{Xmatmom}
\end{equation}
with the convention $\overline J_{-1}=\overline J_{N+1}=0$.
It then follows that
\begin{equation}
\langle n | \widehat X | \omega _k \rangle = \lambda _n \langle n | \omega _k  \rangle
\end{equation}
becomes the difference equation
\begin{equation}
\lambda_n \phi_n(\omega_{k}) = \overline{J}^*_{k} \phi_n(\omega_{k+1}) 
- \overline{B}_{k} \phi_n(\omega_{k}) 
+  \overline{J}_{k-1} \phi_n(\omega_{k-1}) \,, \qquad 0\leq k\leq N \,.
\label{diffphi}
\end{equation}
Equations \eqref{recurphi} and \eqref{diffphi} provide a bispectral problem for 
$\phi_n(\omega_{k})$ which is a discrete version of the bispectral 
problem \eqref{eq:bis} at the root of the previous section.

When $N$ is finite, the couple of operators $\widehat H$ and $\widehat
X$ form by definition a Leonard pair \cite{MR1826654}.  One can deduce
that the eigenvalues $\{\omega_k\}$ of $\widehat H$ are pairwise
distinct and similarly for the eigenvalues $\{\lambda_n\}$ of
$\widehat X$ (see Lemma 1.3.  in \cite{2005JCoAM.178..437T}).
Leonard pairs have been classified \cite{MR1826654} and shown to be in
correspondence with the orthogonal polynomial families of the truncating
part of the Askey tableau.  As a matter of fact, all discrete
hypergeometric polynomials of that scheme, not only the finite
classes, provide admissible $\widehat H$ and $\widehat X$ through
their recurrence relation and difference equation.

Summing up, the Fermionic chains susceptible of admitting a commuting
tridiagonal matrix are those whose specifications are dictated by a
duo of operators $\widehat H$ and $\widehat X$ with the special
properties described above.  Operators that would qualify are for instance two
generators of the Askey-Wilson algebra or, for $q=1$, of the Racah
algebra; these are quadratic algebras which respectively describe the
bispectral properties of the polynomials sitting at the top of the
Askey scheme.  As particular and simpler cases, a moment's thought
will make one realize that two generators of rank-one Lie or
$q$-deformed Lie algebras will meet the requirement that one of these
elements will be represented by an irreducible tridiagonal matrix in
the eigenbasis of the other and vice-versa.  These are the situations
on which we will focus in Sections 7, 8 and 9.

Given such bispectral contexts, the time and band limiting experience
has taught us that nice commuting operators can be simply obtained
from the associated algebraic Heun operator.  This is what we will
explain in the next section before we come to examples.

\section{Algebraic Heun operators and commuting matrices \label{sec:Heunop}}

Looking for a tridiagonal matrix $T$ that commutes with $C$, in the
spirit of Section \ref{sec:BTL}, we introduce the ``discrete -
discrete'' version of the algebraic Heun operator \eqref{D}.  As per
\eqref{eq:heun2}, we take this operator to be
\cite{2018CMaPh.364.1041G} the following bilinear combination of the
two operators that define the bispectral problem:
\begin{equation}
 \widehat T= \{ \widehat X , \widehat H \}  + \tau [ \widehat X , 
 \widehat H] + \mu \widehat X + \nu  \widehat H \,.
 \label{Heun}
\end{equation}
At this point the parameters $\tau, \mu, \nu$
are free.  (Note that allowing for redefinition by an irrelevant
overall factor, the coefficient of $\{ \widehat X , \widehat H \}$ has
been set equal to 1.)
It is immediate to see that $\widehat{T}$ is tridiagonal in both the position basis
\begin{align}
  \widehat T 
  |n\rangle&=J_{n-1}\left(\lambda_{n-1}(1+\tau)+\lambda_n(1-\tau)+\nu 
  \right)|n-1\rangle \nonumber\\
  &+(\mu \lambda_n-2B_n\lambda_n-\nu B_n)|n\rangle\nonumber \\
  &+J_n \left(\lambda_n(1-\tau)+\lambda_{n+1}(1+\tau)+\nu \right)|n+1\rangle  \ ,
 \label{Tpos}
\end{align}
and the momentum basis
\begin{align}
 \widehat T |\omega_k \rangle&=\overline J_{k-1}(\omega_{k-1}(1-\tau)+\omega_k(1+\tau)+\mu)|\omega_{k-1}\rangle \nonumber \\
 &+(\nu \omega_k-2 \overline B_k\omega_k-\mu \overline B_k)|\omega_k\rangle\nonumber \\
 &+\overline J_k(\omega_k(1+\tau)+\omega_{k+1}(1-\tau)+\mu)|\omega_{k+1}\rangle   \ .
\end{align}
As a matter of fact, it has been shown in Ref.  \cite{MR2277640} that
$\widehat{T}$ is the most general operator which is tridiagonal in
both bases in finite-dimensional situations.

Let $\widehat{T}_{mn} = \langle m | \widehat T | n 
\rangle$ and  define the ``chopped'' matrix $T$ by
\begin{equation}
	T=|\widehat T_{mn}|_{0\leq m,n \leq \ell} \,.
	\label{Tmat}
\end{equation}	
Following the results of Refs.  \cite{Perline:1987:DTL:37170.37175,
2018CMaPh.364.1041G}, we know that $T$ and $C$ will commute,
\begin{equation}\label{eq:comTC}
 [T,C]=0 \,,
\end{equation}
if the parameters in $\widehat T$ \eqref{Heun} are given by
\begin{equation}\label{eq:cons1}
    \tau=0\ ,\quad  
	\mu= -(\omega_K+\omega_{K+1}) \quad \text{and} \qquad 
	\nu= -(\lambda_\ell+\lambda_{\ell+1}) \ .
\end{equation}
Indeed, with the particular value of $\nu$ given by \eqref{eq:cons1},
we see that the matrix $\widehat T$ leaves the subspace $\{|n\rangle\,,
n=0,1,\dots, \ell\}$ invariant.  Therefore, $T$ commutes with $\pi_1$.
Similarly, with $\mu$ specified by \eqref{eq:cons1}, $\widehat T$ leaves the subspace
$\{|\omega_k\rangle, k=0,1,\dots, K\}$ invariant and $T$ commutes
with $\pi_2$.  Finally, in view of \eqref{Cpi},
it is easy to to see that \eqref{eq:comTC} holds.

The main result of this section is that the tridiagonal matrix $T$ \eqref{Tmat} 
i.e.
\begin{equation}
 T =\begin{pmatrix}
             d_{0} & t_{0} & \\
             t_{0} & d_{1} & t_{1} \\
             & t_{1} & d_{2} & t_{2} \\
             && \ddots & \ddots & \ddots \\
            &&&t_{\ell-2} & d_{\ell-1} & 
			t_{\ell-1}\\
             &&&& t_{\ell-1} & d_{\ell}
            \end{pmatrix}\,,
			\label{Tmatfinal}
\end{equation}
whose nonzero matrix elements are given by (see \eqref{Tpos}) 
\begin{align}
t_{n} &= J_n(\lambda_n+\lambda_{n+1}-\lambda_\ell-\lambda_{\ell+1}) 
\,, \\
d_{n} &= - B_n (2\lambda_n - \lambda_\ell - \lambda_{\ell+1})
- \lambda_n (\omega_K+\omega_{K+1})\,
\label{Tmatelems}
\end{align}	
commutes with the correlation matrix \eqref{eq:comTC}.  
A key ingredient obviously is the operator
$\widehat X$ defined in \eqref{Xmat}.  In the following sections, we
apply this construction to examples of both finite and semi-infinite
free Fermionic chains.

If $t_n\neq 0$ (which is the case in the examples below), $T$ is
non-degenerate (see e.g. Lemma 3.1 in Ref.  \cite{MR1826654}) and the
commuting matrices $T$ and $C$ have a unique set of common
eigenvectors.  Since $T$ is tridiagonal, its eigenvectors can be
readily computed numerically.  By acting with $C$ on these
eigenvectors, the eigenvalues of $C$ can be easily obtained.  The
eigenvalues of the entanglement Hamiltonian ${\cal H}$, and therefore
the entanglement entropy of the model, can then also be
straightforwardly determined.

\section {The chain based on $\mathfrak{su}(2)$ \label{sec:su2}}

In this section, and the subsequent ones, we use unitary
representations of Lie and $q$-deformed algebras to identify
appropriate pairs of bispectral Hermitian operators $\widehat H$ and
$\widehat X$.  We then construct the Heun operator to obtain explicit
examples of matrices $T$ that commute with the respective entanglement
Hamiltonians.

We begin with the simplest case, that is, $\mathfrak{su}(2)$.
The spin $s$ ($s\in \mathbb{Z}/2$) representation of $\mathfrak{su}(2)$ is given by
\begin{align}
s^{x}&=\frac{1}{2}\sum_{n=0}^{2s} \sqrt{(n+1)(2s-n)} \Big(|n\rangle 
\langle n+1| +|n+1\rangle \langle n|\Big) \,, \\
s^{y}&=-\frac{i}{2}\sum_{n=0}^{2s} \sqrt{(n+1)(2s-n)} \Big(|n\rangle 
\langle n+1| -|n+1\rangle \langle n|\Big) \,, \\
    s^z&=-\sum_{n=0}^{2s} (n-s) |n\rangle \langle n| \,.
\end{align}
We choose 
\begin{equation}
    \widehat H=\cos(\theta) s^z -\sin(\theta) s^x - b \,,
\end{equation}
where $b$ and $\theta$ are real constants. In view of \eqref{eq:Hh}, we study a chain 
with $N=2s$ and with parameters (see \eqref{eq:Hff}) 
given by
\begin{equation}
    B_n=\cos(\theta)(n-s)+b\ , \qquad J_n 
	=-\frac{1}{2}\sin(\theta)\sqrt{(n+1)(2s-n)} \,.
\end{equation}
To diagonalize $\widehat{H}$, 
we observe that $\widehat{H} = U (s^{z} - b) U^{\dagger}$ with $U=e^{i \theta s^{y}}$,
and hence $ \widehat{H} |\omega_k \rangle = \omega_k |\omega_k 
\rangle $ with 
\begin{equation}
     |\omega_k \rangle = U |2s-k \rangle \qquad \text{and} \qquad
	\omega_k = k-s-b \,,  
	\label{eigen}
\end{equation}
where $k=0,1,\dots,2s$. 
The integer $K$ in \eqref{Kdef}
is the unique integer satisfying\footnote{We choose $b$ such that $K\in \{0,1,\dots,N\}$.  The
other case $K<0$ (resp.  $K>N$) corresponds to an empty (resp.  full)
ground state which is not interesting from the point of view of the
entanglement entropy.}  
$s+b-1\leq K < s+b$.
Let us mention that $\phi_n(\omega_k) = \langle n|\omega _k \rangle = \langle n| U |k \rangle$ are
given in terms of the Krawtchouk polynomials in this case.

The operator $\widehat X$ \eqref{Xmat} can be chosen as 
$\widehat X= s^z$, which is diagonal in the position basis with 
$\lambda_n=s-n$.
We observe that 
\begin{equation}
s^{z} = U \left(\cos(\theta) s^{z} + \sin(\theta) s^{x} \right) U^{\dagger} \,.
\end{equation}
Hence, in the momentum basis, in light of the first equation in \eqref{eigen}, $\widehat X$ is 
given by
\begin{align}
    \widehat X&= \cos(\theta) \sum_{k=0}^{2s} (k-s) |\omega_k\rangle \langle \omega_k|\nonumber\\
    &+ \frac{1}{2}\sin(\theta)\sum_{k=0}^{2s} \sqrt{(k+1)(2s-k)} \Big(|\omega_k\rangle \langle \omega_{k+1}| 
	+|\omega_{k+1}\rangle \langle \omega_k|\Big) \,.
\end{align}
Comparing with the general form \eqref{Xmatmom} for $\widehat X$ in the momentum basis, we have
\begin{equation}
   \overline B_k= -\cos(\theta)  (k-s)\ , \qquad \overline J_k 
   =\frac{1}{2}\sin(\theta)\sqrt{(k+1)(2s-k)} \,.
\end{equation}
The Heun operator associated to the Lie algebra $\mathfrak{su}(2)$ has been studied previously in \cite{cramp2019heun}. 
We conclude that the matrix $T$ is given by \eqref{Tmatfinal} with 
\begin{align}
t_n&= \sin(\theta) (n-\ell)\sqrt{(n+1)(2s-n)}
\,,\\
d_{n} &=  \left[\cos(\theta)(n-s)+b\right](2n-2\ell-1) + (s-n)(2s-2K+2b-1)\,.
\end{align}

\section {The chain based on $\mathfrak{su}(1,1)$ \label{sec:su11}}

In this section, we focus on the irreducible discrete series unitary representation 
of the Lie algebra $\mathfrak{su}(1,1)$ given by (see e.g. 
\cite{Wybourne1974})
\begin{align}
\sigma^{x}&=\frac{1}{2}\sum_{n=0}^{\infty} \sqrt{(n+1)(\kappa+n)} 
\Big(|n\rangle \langle n+1| +|n+1\rangle \langle n|\Big) \,, \\
\sigma^{y}&=\frac{i}{2}\sum_{n=0}^{\infty} \sqrt{(n+1)(\kappa+n)} 
\Big(|n\rangle \langle n+1| -|n+1\rangle \langle n|\Big) \,, \\
    \sigma^z&=\sum_{n=0}^{\infty} \left(n+\frac{\kappa}{2}\right) 
	|n\rangle \langle n| \,,
\end{align}
where $\kappa$ is a real positive parameter. Indeed, one can show that
\begin{equation}
	\left[\sigma^{x}\,, \sigma^{y} \right] = -i \sigma^{z}\,, \qquad
	\left[\sigma^{z}\,, \sigma^{x} \right] = i \sigma^{y}\,, \qquad
	\left[\sigma^{z}\,, \sigma^{y} \right] = -i \sigma^{x}\,.
\end{equation}	
We choose for $\widehat{H}$
\begin{equation}
    \widehat H^{ell} = \cosh(\theta) \sigma^z -\sinh(\theta) 
	\sigma^x + b ,\label{eq:hell}
\end{equation}
where $b$ and $\theta$  are real. The superscript ``ell''  stands for elliptic. To justify this name, we recall
that a rotation by an element of the group $SU(1,1)$ of a generic
element $l_x \sigma^x+l_y \sigma^y+l_z \sigma^z$ preserves the
non-definite form $l_x^2+l_y^2 - l_z^2$.  For the Lie element $\cosh(\theta) \sigma^z -\sinh(\theta) 
	\sigma^x $ in
\eqref{eq:hell}, this non-definite form is negative with the element thus belonging to the elliptic orbit.

We are therefore studying in this section a chain with an infinite 
number of sites. In view of \eqref{eq:Hh}, the parameters of the Hamiltonian
$\widehat{\mathcal{H}}$ defined by \eqref{eq:Hff} are given by
\begin{equation}
   B_n^{ell}= -\cosh(\theta)\left(n+\frac{\kappa}{2}\right) - b\ , \quad 
   J_n^{ell} =-\frac{1}{2}\sinh(\theta)\sqrt{(n+1)(\kappa+n)}.
\end{equation}
To obtain the eigenvalues and eigenvectors of $\widehat H^{ell}$ \eqref{eq:hell},
we note here that $\widehat H^{ell} = U (\sigma^{z} + b) U^{\dagger}$ with $U=e^{i \theta \sigma^{y}}$,
and  find that $ \widehat H^{ell} |\omega_k \rangle = \omega_k |\omega_k \rangle $ with
\begin{equation}
     |\omega_k \rangle = U |k \rangle \qquad \text{and} \qquad \omega_k =  k+\frac{\kappa}{2}+b \,. 
	\label{eigen2}
\end{equation}
for $k=0,1,\dots$.
Let us mention that the wavefunctions $\phi_n(\omega_k)$ are 
expressed in terms of the Meixner
polynomials in this case.

The operator $\widehat X$ is  taken to be $\widehat X= \sigma^z$,
and is diagonal in the position basis with $\lambda_n=n+\frac{\kappa}{2}$.
We observe that
\begin{equation}
    \widehat X^{ell} = U \left( \cosh(\theta) \sigma^z +\sinh(\theta) 
	\sigma^x \right) U^{\dagger} \,.
\end{equation}
Proceeding as for the $\mathfrak{su}(2)$ model and referring to
\eqref{Xmatmom}, we observe that the expression of $\widehat X$ in the
momentum basis involves the following coefficients:
\begin{equation}
   \overline B^{ell}_k= -\cosh(\theta)\left(k+\frac{\kappa}{2}\right)\ , \qquad 
   \overline J^{ell}_k =\frac{1}{2}\sinh(\theta)\sqrt{(k+1)(k + \kappa)}.
\end{equation}
The Heun operator associated to the Lie algebra $\mathfrak{su}(1,1)$ has been studied previously in \cite{cramp2019heun}. 
We conclude that the matrix $T$ in this case is given by \eqref{Tmatfinal} with 
\begin{align}
t_n&= -\sinh(\theta) (n-\ell)\sqrt{(n+1)(\kappa + n)}\,,\\
d_{n} &= \left[\cosh(\theta)(n+\frac{\kappa}{2})+b\right](2n-2\ell-1)
-(n+\frac{\kappa}{2})(\kappa+2K+2b+1)
\,.
\end{align}

\section {The chain based on $\mathfrak{so}_q(3)$ at q root of unity \label{sec:o3}}

In this section, we offer a final explicit example based on an
irreducible unitary representation of the $q$-deformed Lie algebra
$\mathfrak{so}_q(3)$ at $q$ root of unity.  Let $N$ be a positive
integer and $d=1,2,\dots N-1$.  There is a $(d+1)\times (d+1)$ irreducible representation of
$\mathfrak{so}_q(3)$ with $q=\exp(2i\pi/N)$ given by
\cite{1996LMaPh..37..173S}
\begin{align}
K_1&=-\frac{1}{2}\sum_{n=0}^{d-1} 
\sqrt 
{\frac{ \sin \left( {\frac {\pi(n+1)}{N}} \right) 
    \sin \left( {\frac {\pi \left( d-n \right) }{N}} \right) }
   { \cos \left( {\frac {\pi \left( d-2n-2 \right) }{2N}} \right) 
     \cos \left( {\frac {\pi \left( d-2n \right) }{2N}} \right)   }}
\Big(|n\rangle \langle n+1| +|n+1\rangle \langle n|\Big) \,, \\
    K_0&=\sum_{n=0}^{d} \sin \left( \frac {\pi\, \left( 2\,n-d \right) }{2N} \right) 
	|n\rangle \langle n| \,.
\end{align}
We define $K_2=e^{i\pi/(2N)} K_0 K_1-e^{-i\pi/(2N)} K_1K_0 $. Then, one gets 
\begin{eqnarray}
&&e^{i\pi/(2N)} K_1 K_2-e^{-i\pi/(2N)} K_2K_1=-\sin^2\left(\frac{\pi}{N}\right) K_0\,,\\
&&e^{i\pi/(2N)} K_2 K_0-e^{-i\pi/(2N)} 
K_0K_2=-\sin^2\left(\frac{\pi}{N}\right) K_1 \,,
\label{qcom}
\end{eqnarray}
thus realizing the defining relations of $\mathfrak{so}_q(3)$ (we have
changed the normalisation of the generators $K_i$ for later
convenience).

We take for $\widehat H$
\begin{equation}
    \widehat H^{\mathfrak{so}} = K_1 + b ,\label{eq:hoq}
\end{equation}
where $b$ is a real constant. 
This defines a chain with $d+1$ sites. In view of \eqref{eq:Hh}, the couplings of the Hamiltonian 
$\widehat{\mathcal{H}}$ defined by \eqref{eq:Hff} are in this case given by
\begin{equation}
   B_n^{\mathfrak{so}}=-b\ , \quad 
   J_n^{\mathfrak{so}} = -\frac{1}{2}
\sqrt 
{\frac{ \sin \left( {\frac {\pi(n+1)}{N}} \right) 
    \sin \left( {\frac {\pi \left( d-n \right) }{N}} \right) }
   { \cos \left( {\frac {\pi \left( d-2n-2 \right) }{2N}} \right) 
     \cos \left( {\frac {\pi \left( d-2n \right) }{2N}} \right)   }}.
\end{equation}
Let us remark that when the number of sites is related to the order of
the unity root, i.e. when $d=N-2$, these reduce to
\begin{equation}
   B_n^{\mathfrak{so}}=-b\ , \quad 
   J_n^{\mathfrak{so}} = -\frac{1}{2}.
\end{equation}
Hence, the model treated here generalizes the homogeneous chain studied in \cite{Cramp__2019}.

Note that the $q$-commutation relations \eqref{qcom} of
$\mathfrak{so}_q(3)$ are symmetric under the exchange $K_0 \leftrightarrow
K_1$; hence, in the present representation where this
permutation is unitarily realized, $K_1$ has the same spectrum
as $K_0$ \cite{1996LMaPh..37..173S}.
Therefore, $\widehat H$ given by \eqref{eq:hoq} is diagonalized as follows: for $k=0,1,\dots d$,
\begin{equation}
    \widehat{H}^{\mathfrak{so}} |\omega_k \rangle 
	=\omega_k |\omega_k \rangle \,, \qquad \omega_k = \sin \left( \frac {\pi\, \left( 2\,k-d \right) }{2N} \right)  +b \,.
\end{equation}
Let us mention that the wavefunctions $\phi_n(\omega_k)$ involve the
$q$-ultraspherical polynomials at $q$ a root of
unity.  It is interesting to realize that the finite Chebychev
polynomials that occur in the uniform chain are a special case of
these $q$-polynomials.

The operator $\widehat X$ can be chosen as $\widehat X^{\mathfrak{so}}= K_0$,
which is diagonal in the position basis with 
$\lambda_n=  \sin \left( \frac {\pi \left( 2n-d \right) }{2N} \right) $. 
In the momentum basis, this operator $\widehat X$ is also tridiagonal and reads
\begin{equation}
   \overline B^{\mathfrak{so}}_k=0\ , \qquad 
   \overline J^{\mathfrak{so}}_k =\frac{1}{2}\sqrt 
{\frac{ \sin \left( {\frac {\pi(k+1)}{N}} \right) 
    \sin \left( {\frac {\pi \left( d-k \right) }{N}} \right) }
   { \cos \left( {\frac {\pi \left( d-2k-2 \right) }{2N}} \right) 
     \cos \left( {\frac {\pi \left( d-2k \right) }{2N}} \right)   }}.
\end{equation}
We conclude that the matrix $T$ is given by \eqref{Tmatfinal} with 
\begin{align}
t_n=& 2\cos\left(\frac{\pi}{2N}\right)\sin\left(\frac{\pi(\ell-n)}{2N} \right)
\cos\left(\frac{\pi(\ell+n-d+1)}{2N} \right)\nonumber\\
&\times 
\sqrt 
{\frac{ \sin \left( {\frac {\pi(n+1)}{N}} \right) 
    \sin \left( {\frac {\pi \left( d-n \right) }{N}} \right) }
   { \cos \left( {\frac {\pi \left( d-2n-2 \right) }{2N}} \right) 
     \cos \left( {\frac {\pi \left( d-2n \right) }{2N}} \right)   }} 
	 \,, \\
d_{n}=& -2\cos\left(\frac{\pi}{2N}\right)\bigg[ 
b\sin\left(\frac{\pi(2\ell-d+1)}{2N}  \right) \nonumber\\
&\qquad + \sin\left(\frac{\pi(2n-d)}{2N}  
\right)\sin\left(\frac{\pi(2K-d+1)}{2N}  \right) \bigg]\,.
\end{align}
This coincides with the matrix found in \cite{{2018arXiv180500078E}} and \cite{{Cramp__2019}} when $d=N-2$.


\section{Concluding remarks}

This paper has discussed entanglement in free Fermionic chains and
focused in particular on the challenges associated to the
diagonalization of the entanglement Hamiltonian.  It has underscored
in this respect the connection that these studies bear with the
classic treatment of time and band limiting in signal processing.
This article has illustrated how the methods developed in the latter
context can be usefully imported in the entanglement analyses of
Fermionic chains.  The key feature that has thus been adapted is the
existence of a second order differential (or difference) operator that
commutes with the non-local limiting operator.  In time this
remarkable fact has been understood to arise from an underlying
bispectral situation, and recently \cite{2018CMaPh.364.1041G} the 
related algebraic Heun operator construct was seen to lead to these commuting
operators.  This was reviewed here and was seen to be transposable to
the entanglement of Fermionic chains.

The specifications of chains which have a bispectral underpinning have
been characterized.  Involved are two operators ($\widehat H$ and
$\widehat X$) which are diagonal in the momentum and position bases
respectively and tridiagonal in the other.  They define the bispectral
problem that the wavefunctions satisfy.  Attached to chains of that
type are algebraic Heun operators that readily yield a tridiagonal
matrix that commutes with the restricted correlation matrix which is
the fundamental operator that needs to be diagonalized.  It was
pointed out that the bispectral operators generate algebraic structures
of interest and are connected to orthogonal polynomials.  With that
perspective, three pairs of bispectral operators were identified from
representations of the Lie and q-deformed algebras $\mathfrak{su}(2)$,
$\mathfrak{su}(1,1)$ and $\mathfrak{so}_q(3)$.  The corresponding free
Fermionic chains were introduced and the commuting matrices presented.
The first model gave an example of a finite chain, the second of a
semi-infinite one and the third based on representations of
$\mathfrak{so}_q(3)$ at $q$ a root of unity offered a one-parameter
generalization of the chain with uniform couplings.

A number of interesting questions are pending and deserve further
investigations.  In all our considerations, the bipartition of the
chains has been defined by considering one part as the subset of sites
consisting of consecutive nodes starting with the first one.  It would
obviously be of relevance to extend the approach to other space
limiting.  Studies of entanglement of Fermions (and Bosons) on
different graphs have been undertaken \cite{refId0, Jafarizadeh_2018}.
We plan on examining how the considerations developed in this paper
could extend in that context.  It would also be nice to carry this out
in field theory especially in the Schrodinger representation (see in 
particular \cite{Callan:1994py}) that
Roman Jackiw has at times advocated \cite{Floreanini:1986tq, Floreanini:1987gr}.

\paragraph{Acknowledgements} 
The authors are grateful to the editors for the invitation to
contribute to this volume in honour of Roman Jackiw.  N. Cramp\'e
warmly thanks the Centre de Recherches Math\'ematiques (CRM) for
hospitality and support during his visit to Montreal in the course of
this investigation.  The research of L. Vinet is supported in part by
a Discovery Grant from the Natural Science and Engineering Research
Council (NSERC) of Canada.
 

\begin{thebibliography}{10}

\bibitem{2004JSMTE..06..004P}
I.~{Peschel}, ``{On the reduced density matrix for a chain of free
  electrons},'' \href{http://dx.doi.org/10.1088/1742-5468/2004/06/P06004}{{\em
  Journal of Statistical Mechanics: Theory and Experiment} {\bfseries 2004}
  no.~6, (Jun, 2004) 06004},
  \href{http://arxiv.org/abs/cond-mat/0403048}{{\ttfamily
  arXiv:cond-mat/0403048 [cond-mat.stat-mech]}}.

\bibitem{2018arXiv180500078E}
V.~{Eisler} and I.~{Peschel}, ``{Properties of the entanglement Hamiltonian for
  finite free-fermion chains},'' {\em Journal of Statistical Mechanics: Theory
  and Experiment} {\bfseries 2018} no.~10, (Oct, 2018) 104001,
  \href{http://arxiv.org/abs/1805.00078}{{\ttfamily arXiv:1805.00078
  [cond-mat.stat-mech]}}.

\bibitem{2006PhRvL..96j0503G}
D.~{Gioev} and I.~{Klich}, ``{Entanglement Entropy of Fermions in Any Dimension
  and the Widom Conjecture},''
  \href{http://dx.doi.org/10.1103/PhysRevLett.96.100503}{{\em Phys Rev Lett}
  {\bfseries 96} no.~10, (Mar, 2006) 100503},
  \href{http://arxiv.org/abs/quant-ph/0504151}{{\ttfamily
  arXiv:quant-ph/0504151 [quant-ph]}}.

\bibitem{2013JSMTE..04..028E}
V.~{Eisler} and I.~{Peschel}, ``{Free-fermion entanglement and spheroidal
  functions},'' \href{http://dx.doi.org/10.1088/1742-5468/2013/04/P04028}{{\em
  Journal of Statistical Mechanics: Theory and Experiment} {\bfseries 2013}
  no.~4, (Apr, 2013) 04028}, \href{http://arxiv.org/abs/1302.2239}{{\ttfamily
  arXiv:1302.2239 [cond-mat.stat-mech]}}.

\bibitem{MR710468}
D.~Slepian, ``Some comments on {F}ourier analysis, uncertainty and modeling,''
  \href{http://dx.doi.org/10.1137/1025078}{{\em SIAM Rev.} {\bfseries 25}
  no.~3, (1983) 379--393}. \url{https://doi.org/10.1137/1025078}.

\bibitem{Landau1985}
H.~Landau, {\em An Overview of Time and Frequency Limiting},
  \href{http://dx.doi.org/10.1007/978-1-4613-2525-3_12}{pp.~201--220}.
\newblock Springer US, Boston, MA, 1985.
\newblock \url{https://doi.org/10.1007/978-1-4613-2525-3_12}.

\bibitem{Slepian1961}
D.~Slepian and H.~O. Pollak, ``{Prolate spheroidal wave functions, Fourier
  analysis and uncertainty. I},'' {\em The Bell System Technical Journal}
  {\bfseries 40} no.~1, (1961) 43--63.

\bibitem{1986CMaPh.103..177D}
J.~J. {Duistermaat} and F.~A. {Gr{\"u}nbaum}, ``{Differential equations in the
  spectral parameter},'' \href{http://dx.doi.org/10.1007/BF01206937}{{\em
  Communications in Mathematical Physics} {\bfseries 103} no.~2, (Jun, 1986)
  177--240}.

\bibitem{doi:10.1002/cpa.3160470305}
F.~A. Grünbaum, ``Time-band limiting and the bispectral problem,''
  \href{http://dx.doi.org/10.1002/cpa.3160470305}{{\em Communications on Pure
  and Applied Mathematics} {\bfseries 47} no.~3, (1994) 307--328},
  \href{http://arxiv.org/abs/https://onlinelibrary.wiley.com/doi/pdf/10.1002/cpa.3160470305}{{\ttfamily
  https://onlinelibrary.wiley.com/doi/pdf/10.1002/cpa.3160470305}}.
  \url{https://onlinelibrary.wiley.com/doi/abs/10.1002/cpa.3160470305}.

\bibitem{Koekoek}
R.~Koekoek and R.~F. Swarttouw, ``{The Askey-scheme of hypergeometric
  orthogonal polynomials and its q-analogue},'' Tech. Rep. 98-17, Delft
  University of Technology, Faculty of Information Technology and Systems,
  Department of Technical Mathematics and Informatics, 1998.
\newblock \url{https://homepage.tudelft.nl/11r49/askey.html}.

\bibitem{Koekoek2010}
R.~Koekoek, P.~A. Lesky, and R.~F. Swarttouw,
  \href{http://dx.doi.org/10.1007/978-3-642-05014-5}{{\em Hypergeometric
  Orthogonal Polynomials and Their q-Analogues}}.
\newblock Springer-Verlag, 2010.
\newblock \url{https://doi.org/10.1007/978-3-642-05014-5}.

\bibitem{2017JMP....58c1703G}
F.~A. {Gr{\"u}nbaum}, L.~{Vinet}, and A.~{Zhedanov}, ``{Tridiagonalization and
  the Heun equation},'' \href{http://dx.doi.org/10.1063/1.4977828}{{\em Journal
  of Mathematical Physics} {\bfseries 58} no.~3, (Mar, 2017) 031703},
  \href{http://arxiv.org/abs/1602.04840}{{\ttfamily arXiv:1602.04840
  [math-ph]}}.

\bibitem{Cramp__2019}
N.~Cramp\'e, R.~I. Nepomechie, and L.~Vinet, ``Free-fermion entanglement and
  orthogonal polynomials,''
  \href{http://dx.doi.org/10.1088/1742-5468/ab3787}{{\em Journal of Statistical
  Mechanics: Theory and Experiment} {\bfseries 2019} no.~9, (Sep, 2019)
  093101}. \url{http://dx.doi.org/10.1088/1742-5468/ab3787}.

\bibitem{2003JPhA...36L.205P}
I.~{Peschel}, ``{Letter to the Editor: Calculation of reduced density matrices
  from correlation functions},''
  \href{http://dx.doi.org/10.1088/0305-4470/36/14/101}{{\em Journal of Physics
  A Mathematical General} {\bfseries 36} no.~14, (Apr, 2003) L205--L208},
  \href{http://arxiv.org/abs/cond-mat/0212631}{{\ttfamily
  arXiv:cond-mat/0212631 [cond-mat]}}.

\bibitem{2009JPhA...42X4003P}
I.~{Peschel} and V.~{Eisler}, ``{Reduced density matrices and entanglement
  entropy in free lattice models},''
  \href{http://dx.doi.org/10.1088/1751-8113/42/50/504003}{{\em Journal of
  Physics A Mathematical General} {\bfseries 42} no.~50, (Dec, 2009) 504003},
  \href{http://arxiv.org/abs/0906.1663}{{\ttfamily arXiv:0906.1663
  [cond-mat.stat-mech]}}.

\bibitem{Lee:2014nra}
C.~H. Lee, P.~Ye, and X.-L. Qi, ``{Position-momentum duality in the
  entanglement spectrum of free fermions},''
  \href{http://dx.doi.org/10.1088/1742-5468/2014/10/P10023}{{\em J. Stat.
  Mech.} {\bfseries 1410} no.~10, (2014) P10023},
\href{http://arxiv.org/abs/1403.1039}{{\ttfamily arXiv:1403.1039
  [cond-mat.str-el]}}.

\bibitem{2012PhRvB..86x5109H}
Z.~{Huang} and D.~P. {Arovas}, ``{Entanglement spectrum and Wannier center flow
  of the Hofstadter problem},''
  \href{http://dx.doi.org/10.1103/PhysRevB.86.245109}{{\em Physical Review B}
  {\bfseries 86} no.~24, (Dec, 2012) 245109},
  \href{http://arxiv.org/abs/1201.0733}{{\ttfamily arXiv:1201.0733
  [cond-mat.stat-mech]}}.

\bibitem{2018CMaPh.364.1041G}
F.~A. {Gr{\"u}nbaum}, L.~{Vinet}, and A.~{Zhedanov}, ``{Algebraic Heun Operator
  and Band-Time Limiting},''
  \href{http://dx.doi.org/10.1007/s00220-018-3190-0}{{\em Communications in
  Mathematical Physics} {\bfseries 364} no.~3, (Dec, 2018) 1041--1068},
  \href{http://arxiv.org/abs/1711.07862}{{\ttfamily arXiv:1711.07862
  [math-ph]}}.

\bibitem{MR1826654}
P.~Terwilliger, ``Two linear transformations each tridiagonal with respect to
  an eigenbasis of the other,''
  \href{http://dx.doi.org/10.1016/S0024-3795(01)00242-7}{{\em Linear Algebra
  Appl.} {\bfseries 330} no.~1-3, (2001) 149--203}.
  \url{https://doi.org/10.1016/S0024-3795(01)00242-7}.

\bibitem{2005JCoAM.178..437T}
P.~{Terwilliger}, ``{Two linear transformations each tridiagonal with respect
  to an eigenbasis of the other: comments on the split decomposition},'' {\em
  Journal of Computational and Applied Mathematics} {\bfseries 178} (Jun, 2005)
  437--452, \href{http://arxiv.org/abs/math/0406555}{{\ttfamily
  arXiv:math/0406555 [math.RA]}}.

\bibitem{MR2277640}
K.~Nomura and P.~Terwilliger, ``Linear transformations that are tridiagonal
  with respect to both eigenbases of a {L}eonard pair,''
  \href{http://dx.doi.org/10.1016/j.laa.2006.07.004}{{\em Linear Algebra Appl.}
  {\bfseries 420} no.~1, (2007) 198--207},
  \href{http://arxiv.org/abs/0605316}{{\ttfamily arXiv:0605316 [math.RA]}}.
  \url{https://doi.org/10.1016/j.laa.2006.07.004}.

\bibitem{Perline:1987:DTL:37170.37175}
R.~K. Perline, ``Discrete time-band limiting operators and commuting
  tridiagonal matrices,'' \href{http://dx.doi.org/10.1137/0608016}{{\em SIAM J.
  Algebraic Discrete Methods} {\bfseries 8} no.~2, (Apr., 1987) 192--195}.
  \url{http://dx.doi.org/10.1137/0608016}.

\bibitem{cramp2019heun}
N.~Cramp\'e, L.~Vinet, and A.~Zhedanov, ``{Heun algebras of Lie type},''
  \href{http://dx.doi.org/https://doi.org/10.1090/proc/14788}{{\em Proc. Amer.
  Math. Soc.} (2019) }, \href{http://arxiv.org/abs/1904.10643}{{\ttfamily
  arXiv:1904.10643 [math.RA]}}.

\bibitem{Wybourne1974}
B.~G. Wybourne, {\em {Classical Groups for Physicists}}.
\newblock Wiley, 1974.

\bibitem{1996LMaPh..37..173S}
V.~{Spiridonov} and A.~{Zhedanov}, ``{q-Ultraspherical polynomials for q a root
  of unity},'' \href{http://dx.doi.org/10.1007/BF00416020}{{\em Letters in
  Mathematical Physics} {\bfseries 37} no.~2, (Jun, 1996) 173--180},
  \href{http://arxiv.org/abs/q-alg/9605033}{{\ttfamily arXiv:q-alg/9605033
  [math.QA]}}.

\bibitem{refId0}
M.~A. Jafarizadeh, F.~Eghbalifam, and S.~Nami, ``Entanglement entropy of free
  fermions on directed graphs,''
  \href{http://dx.doi.org/10.1140/epjp/i2017-11805-1}{{\em Eur. Phys. J. Plus}
  {\bfseries 132} no.~12, (2017) 539}.
  \url{https://doi.org/10.1140/epjp/i2017-11805-1}.

\bibitem{Jafarizadeh_2018}
M.~A. Jafarizadeh, F.~Eghbalifam, and S.~Nami, ``Entanglement entropy in the
  spinless free fermion model and its application to the graph isomorphism
  problem,'' \href{http://dx.doi.org/10.1088/1751-8121/aaa1da}{{\em Journal of
  Physics A: Mathematical and Theoretical} {\bfseries 51} no.~7, (Jan, 2018)
  075304}. \url{https://doi.org/10.1088%2F1751-8121%2Faaa1da}.

\bibitem{Callan:1994py}
C.~G. Callan, Jr. and F.~Wilczek, ``{On geometric entropy},''
  \href{http://dx.doi.org/10.1016/0370-2693(94)91007-3}{{\em Phys. Lett.}
  {\bfseries B333} (1994) 55--61},
\href{http://arxiv.org/abs/hep-th/9401072}{{\ttfamily arXiv:hep-th/9401072
  [hep-th]}}.

\bibitem{Floreanini:1986tq}
R.~Floreanini, C.~T. Hill, and R.~Jackiw, ``{Functional Representation for the
  Isometries of de Sitter Space},''
\href{http://dx.doi.org/10.1016/0003-4916(87)90213-2}{{\em Annals Phys.}
  {\bfseries 175} (1987) 345}.

\bibitem{Floreanini:1987gr}
R.~Floreanini and R.~Jackiw, ``{Functional Representation for Fermionic Quantum
  Fields},''
\href{http://dx.doi.org/10.1103/PhysRevD.37.2206}{{\em Phys. Rev.} {\bfseries
  D37} (1988) 2206}.

\end{thebibliography}

\providecommand{\href}[2]{#2}\begingroup\raggedright\endgroup

\end{document}